\documentclass[12pt]{article}
\usepackage{amssymb,amsmath}
\usepackage{epsfig}
\usepackage{graphicx}

\def\shiftdown#1{#1\llap{\lower.04ex\hbox{#1}}}

\begin{document}
\begin{center}
{\Large\bf \boldmath Atomic Cascade in Muonic and
Hadronic Hydrogen Atoms} 

\vspace*{6mm}
{T.S. Jensen, V.P. Popov and V.N. Pomerantsev }\\      
{\small \it Institute of Nuclear Physics, Moscow State University,
119992 Moscow, Russia$^a$ Institution1 }      
\end{center}

\vspace*{6mm}

\begin{abstract}
     The atomic cascade in $\mu^- p$ and $\pi^- p$ atoms has
     been studied with the improved version of the extended cascade model
     in which new quantum mechanical calculations of the differential
     and integral cross sections of the elastic scattering, Stark
     transitions and Coulomb de-excitation have been included for
     the principal quantum number values $n\leqslant 8$ and the
     relative energies $E \geqslant 0.01$ eV. The $X$-ray yields
     and kinetic energy distributions are compared with the
     experimental data.
\end{abstract}
\section {Introduction}
The exotic hydrogen-like atoms are formed in highly excited
states, when a heavy negative particle $(\mu^-, \pi^-,
\mathrm{K}^-, etc.)$ slows down and captures in hydrogen media due
to inelastic collisions. The formation is followed by the atomic
cascade in which
 the initial distributions in the quantum numbers ($n, l$) and
kinetic energy of the exotic atom change due to various processes:
radiative transitions, external Auger effect, elastic scattering,
Stark transitions, Coulomb de-excitation, weak decay and nuclear
absorption (in case of hadronic atoms). A good understanding of
the kinetics of atomic cascade in hydrogen-like exotic atoms is
very important both for the planning and interpretation of the
experiments (see for a review \cite{1}). The first theoretical
study of the atomic cascade was performed more than forty five
years ago by Leon and Bethe \cite{2}. In this paper, and later in
more refined models \cite{3,4}, the rates of the collisional
processes were calculated in the semiclassical approximation at a
fixed kinetic energy ($\sim 1$ eV) and used for the simulation of
the atomic cascade. A number of experiments in which the energy
distributions of $\mu^-p$, $\mu^-d$ and $\pi^-p$ atoms were
measured using various time-of-flight methods \cite{5,6,7,8,9}
showed that the kinetic energy changes during the cascade. In
particular, the existence of high energy components has been
established in pionic hydrogen in the neutron time-of-flight
experiment \cite{7} and in muonic hydrogen in diffusion
experiments \cite{8, 9}. This experiments require a more
sophisticated approach taking the evolution of the kinetic energy
distribution into account during the whole atomic cascade.
Beginning with Markushin \cite{10} recent cascade models take into
account the acceleration and deceleration processes. The extended
standard cascade model (ESCM) \cite{11} introduces a number of
improvements compared to the earlier models: for example, the
scattering from molecular hydrogen at high n is calculated as
opposed to the phenomenological treatment in other cascade models.

   In the recent papers \cite{12,13}, the dynamics of the collisions
   of the excited exotic atoms with hydrogen ones has been studied
   in the framework of the close-coupling approach. The elastic scattering,
   Stark transitions and Coulomb de-excitation were described by a unified
   manner in this approach. The differential and integral
   cross-sections were calculated in a wide range of the principle
   quantum number values and relative energies  for muonic, pionic and
   antiprotonic hydrogen atoms. The vacuum polarization  and
   strong interaction shifts (for hadronic atoms) were taken into
   account (see also this issue).\\
   In the present paper the first theoretical results
   of the X-ray yields and kinetic energy distributions for $\mu^- p$
   and $\pi^- p$ atoms calculated in the new version of the ESCM
   are introduced.
   The new results for the collisional processes
   (\cite{12,13}, see also this issue), especially for Coulomb
   de-excitation process, are used in the present ESCM and lead
   to an essential improvement over the previous cascade calculations.

\section {The extended standard cascade model}

     The cascade in exotic atoms (as in the previous ESCM) is
     divided into the classical domain for high $n$ and the
     quantum mechanical domain for low $n$. In the classical
     domain (for muonic and pionic hydrogen we use $n \geq 9$)
     the classical-trajectory Monte Carlo calculations have been
     included in the cascade model (see for
     details \cite{11}) with the molecular structure of the
     target taken into account for the description of the elastic
     scattering, Stark mixing and Coulomb de-excitation
     processes. The external Auger effect was calculated in the
     semiclassical approximation through  the whole cascade.
     In the upper stage of the cascade
     the distributions in quantum numbers and kinetic energy are
     calculated. Further, these
     distributions were used as input data for the next stage of
     the atomic cascade.
     In the quantum mechanical domain (at $n\leq 8$ )
     the differential and integral cross sections
     for  all $nl\mapsto n'l'$ transitions
     (besides the Auger and radiative de-excitations)
     were calculated in the fully quantum mechanical
     close-coupling approach (see \cite{12,13} and references
     therein) for the initial states with $n\leq 8$ and relative
     energies $E\geq 0.01$ eV. The corresponding arrays of the cross sections
     are included  in the cascade code. Thus the present cascade code
     does not employ any fitting parameters and the
     kinetics of the atomic cascade is treated more accurately.\\
The initial conditions for the cascade calculations are defined by
the initial distributions in the quantum numbers $n$ and $l$ and
the laboratory kinetic energy of the exotic atoms. In the simplest
picture, the exotic particle is captured by the proton of the
hydrogen atom in a state with $n_i \sim \sqrt{M}$ (M is a reduced
mass of exotic atom in atomic units). More elaborate approaches
\cite{14,15}  take the molecular effects of the target into
account and predict distribution in the initial $n_i$ with the
peak shifted towards lower values and non-statistical
$l_i$-distribution. In the present cascade calculations the
influence of the initial parameters are also studied. In order to
obtain good statistics the cascade calculations were performed
with $10^6$ events.

   \section {Results}
   \subsection {Muonic hydrogen}
   Muonic hydrogen atom is the simplest of exotic atoms and its
   experimental and theoretical studies are the best probe
   for the investigation of the
   various processes in exotic atom from hydrogen atom or molecule
   collisions.
   The present cascade calculations in muonic hydrogen have been done in
   the density range ($10^{-7}-1$) (in units of Liquid Hydrogen
   Density,
   $LHD = 4.25\cdot 10^{22} atoms/cm^3$) using various initial conditions.
   The results are
   compared with the experimental data \cite{16,17,18} for relative $K$ $X$-ray yields.
   The energy distributions of muonic hydrogen and Doppler broadening
   of the X-ray lines are also introduced.\\

The dependence of the calculated absolute $K$  $X$-ray yields on
hydrogen density for the $\mu^- p$ atom are shown in Figure 1.

\begin{figure}[h!]
     \centerline{\includegraphics[width=0.7\textwidth,keepaspectratio]{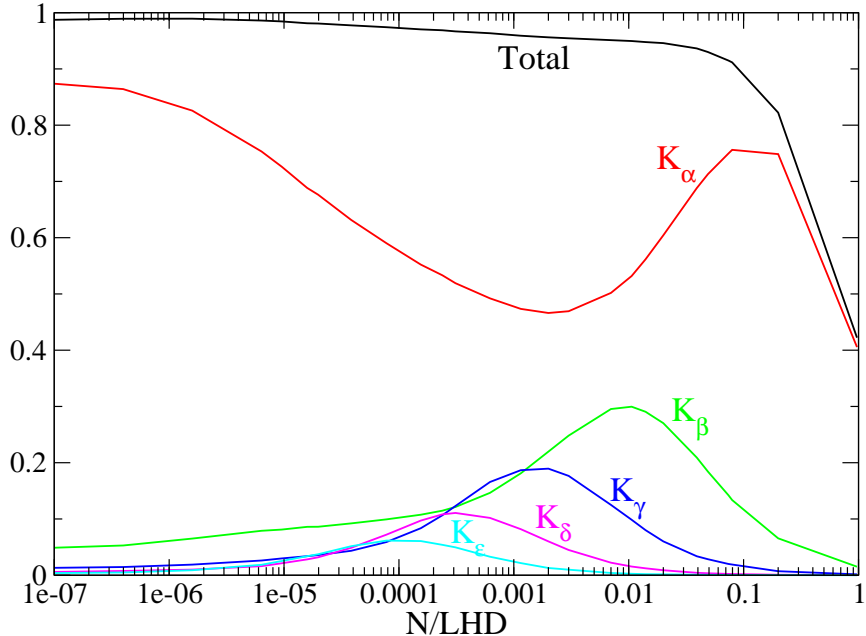}}
     \caption{The density dependence of the absolute $K$ $X$-ray
      yields in muonic hydrogen.}
     \end{figure}

 In contrast to the practically linear dependence of the absolute
$K_{tot}$ yield on the density in the wide range from $10^{-7}$
LHD  to $\sim 10^{-2}$ LHD the individual $K_i$ yields ($i =
\alpha, \beta,$ etc.) have a more complicated density dependence
due to the competition between the various collisional and
non-collisional processes during the cascade. The atomic cascade
at low target densities is mainly dominated by the radiative
de-excitation proceeding through the circular states that result
in $2p\rightarrow 1s$ transition whereas the transitions
$np\rightarrow 1s$ are much weaker. We see the corresponding picture
in Figure 1. At very low density the kinetic energy
distribution of the muonic atom must conserve it initial kinetic
energy distribution. While the density is increased the role of
the collisional processes enhances leading to both the decrease of
$K_\alpha$ yield and simultaneously the increase of the other $K_i$
yields ($i\neq \alpha$), conserving the total absolute yield
practically unchangeable. With increasing density the Coulomb
de-excitation becomes also more efficient (especially at low
energies) and leads to the acceleration of the exotic atom and higher
populations of the $np$ states with $n > 2$ due to the Stark
transitions. At the target densities more than $\sim 10^{-3}\div
10^{-2}$ LHD the collisional de-excitation $n\rightarrow n'
(n'<n)$ begins to dominate the radiative transitions (external Auger
effect and Coulomb transitions) resulting in the decrease of the
$K_{\geq\beta}$ yields. Since the collisional de-excitations to
the $1s$ state of the muonic hydrogen are strongly suppressed
(more or about $2$ keV energy is released) we observe the
enhancement of the absolute $K_{\alpha}$ yield due to the
collisional transitions to $2p$ state and the Stark $2s\rightarrow
2p$ transition at the energy above the threshold. Thus the swift
suppression of the absolute $K_\alpha$ yield at the densities from
$10^{-7}$ LHD  to $\sim 10^{-3}$ LHD  and increase at more higher
densities can be explained by the competition of the different
processes during the atomic cascade in which the evolution of the
kinetic energy is
taken into account through the whole cascade.\\

        The discussed above regularities in the absolute yields can be
justified or vice versa in comparison of the calculated relative
X-ray yields for the $\mu^- p$ with the experimental data
\cite{16,17,18} introduced in Figure 2.

     \begin{figure}[h!]
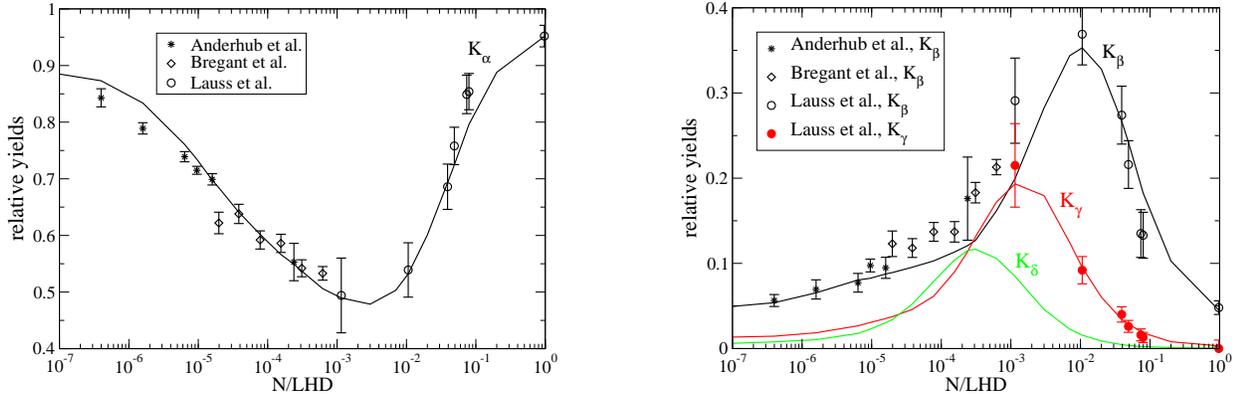

     \centerline{\includegraphics[width=0.45\textwidth,keepaspectratio]{k_a_mu.eps}
     \hfill \includegraphics[width=0.45\textwidth,keepaspectratio]{k_b_mu.eps}}
     \caption{The density dependence of the relative $X$-ray yields
      in muonic hydrogen: $K_\alpha$ on the left; $K_\beta$, $K_\gamma$
      and $K_\delta$ yields (black, red and green, respectively) on the right.
      The experimental data are from \cite{16,17,18}.}
      \end{figure}
The agreement between theoretical results and experimental data is
very good practically for all densities under consideration. The
observable disagreements can be a subject for further both
theoretical and experimental studies. In our opinion, there are
also a number of problems which demand more elaborate
experimental studies: a proper separation of the different $k$
lines, inclusion a Doppler broadening effect in the analysis of
the experimental data and so on. It would be very important to
check the present results directly by measuring the absolute
$X$-ray yields at some values of the target density.\\

The kinetic energy distribution of the exotic atom changes during
the cascade and is a more refined probe of the theoretical
approaches to the description of the cascade processes. Using the
cascade model described above we calculated the $\mu^- p$ kinetic
energy distribution at the moment of the $K_\alpha, K_\beta,$ etc.
radiative transitions and the corresponding Doppler broadening of
the $1s$ line due to the kinetic energy distribution of the exotic
atom at the instant of the $np\rightarrow 1s$ radiative
transitions ($n=2\div 4$) at the target pressure $10$ bar ($\sim
10^{-2}$ LHD). The results are shown in Figure 3.

    \begin{figure}[h!]
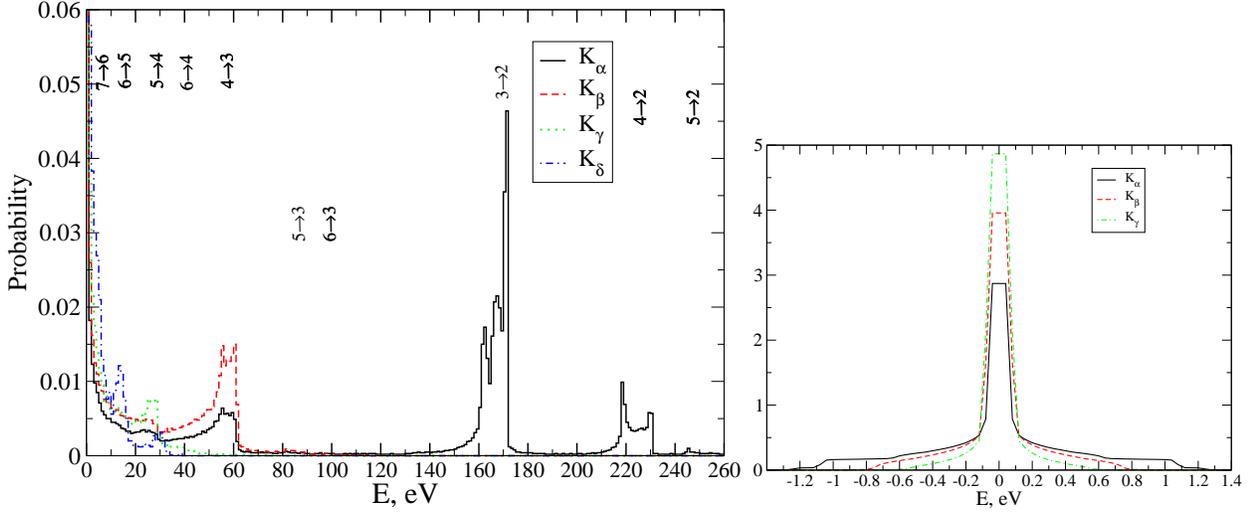

    \centerline{
    \includegraphics[width=0.6\textwidth,keepaspectratio]{dist_mu_10b.eps}
    \hfill  \includegraphics[width=0.4\textwidth,keepaspectratio]{dopp.eps}}
    \caption{On the left: the kinetic-energy distribution of $\mu^- p$ at the instant
    of the radiative $np \rightarrow 1s$ transitions in gaseous hydrogen
    at pressure $10$ bar. On the right: the $K$-line shapes after Doppler broadening
    due to the kinetic-energy distribution.}
    \end{figure}

 As it is seen
from Figure 3, the kinetic energy distribution has distinctive
high energy structures arising from different Coulomb
transitions with $\Delta n\geq 1$ preceding the radiative
de-excitation. The complicated shapes of these structures can be
simply explained by the interplay of three factors: the kinetic energy
distribution of exotic atom before Coulomb transition, the
anisotropy of the angular distribution in the Coulomb
de-excitation process and, finally, the deceleration due to the
elastic scattering and Stark transitions after Coulomb
de-excitation but before the radiative de-excitation. \\
 The energy distribution at low n can be determined by measuring
 the Doppler broadening of $K$ $X$-ray lines. As demonstrated in
 Figure 3, the high-energy components lead to the significant
 Doppler broadening, which especially pronounced for $K_\alpha$
 line. The resulting shapes  and  widths for $K_\alpha, K_\beta$ and $K_\gamma$
 lines at target pressure $10$ bar are
 shown in Fig. 3 (on the right panel).  A significant spreading up to
 $\pm 1.25$ eV for $K_\alpha$ line was found.

   The essential acceleration during cascade is illustrated in
   Figure 4. Here the calculated density dependence of the mean
   kinetic energy at the instant of the radiative $np\rightarrow
   1s$ de-excitation is shown.
     \begin{figure}[h!]
     \centerline{
     \includegraphics[width=0.7\textwidth,keepaspectratio]{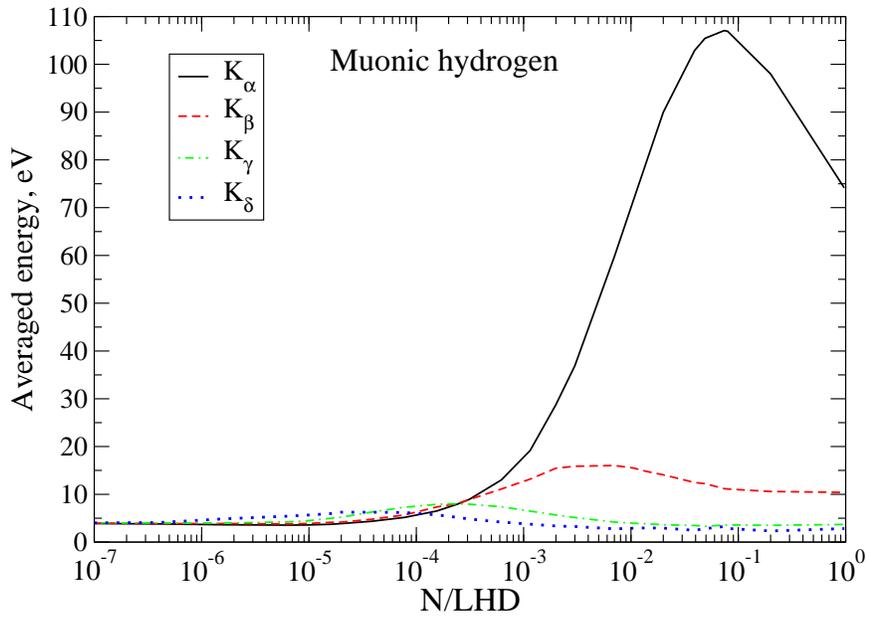}}
     \caption{ The density dependence of the mean kinetic energy of the $\mu^- p$ at the
     instant of the different radiative $np\rightarrow 1s$ transitions.}
     \end{figure}
At the densities from $10^{-7}$ LHD to $\sim 2\cdot 10^{-4}$ LHD
the mean energy slowly increases for all lines and changes
correspondingly from $4$ eV to $8$ eV. At more higher densities
the mean energy of the $K_\beta$ line reaches the maximum value
$\sim 20$ eV  in the density interval $10^{-3}-10^{-2}$ LHD. The
most pronounced increase of the mean energy was obtained for
$K_\alpha$ line with the maximum more than $100$ eV at the density
$\sim 8\cdot 10^{-2}$ LHD.

               \subsection {Pionic hydrogen}
The present study of the atomic cascade in pionic hydrogen was
focused on the experimental data for the absolute $X$ ray yields
\cite{19} and the neutron time-of-flight spectrum \cite{7}
obtained at LHD. The results of our cascade calculations for the
absolute $X$-ray yields as a function of the target density are
shown in Figure 5 in comparison with the experimental data
\cite{19}. The nuclear absorption in the hadronic atoms results to
a strong suppression of the $K X$-ray yields, beginning from very
low densities. As a whole the calculated yields are in a good
agreement with the experimental data. In the calculation we
introduced the additional parameter simulating the induced
absorption from $np$ states. The value of this parameter used in all present
calculations for $2p$-state was chosen equal to $0.045$ meV.

     \begin{figure}[h!]
     \centerline{
     \includegraphics[width=0.7\textwidth,keepaspectratio]{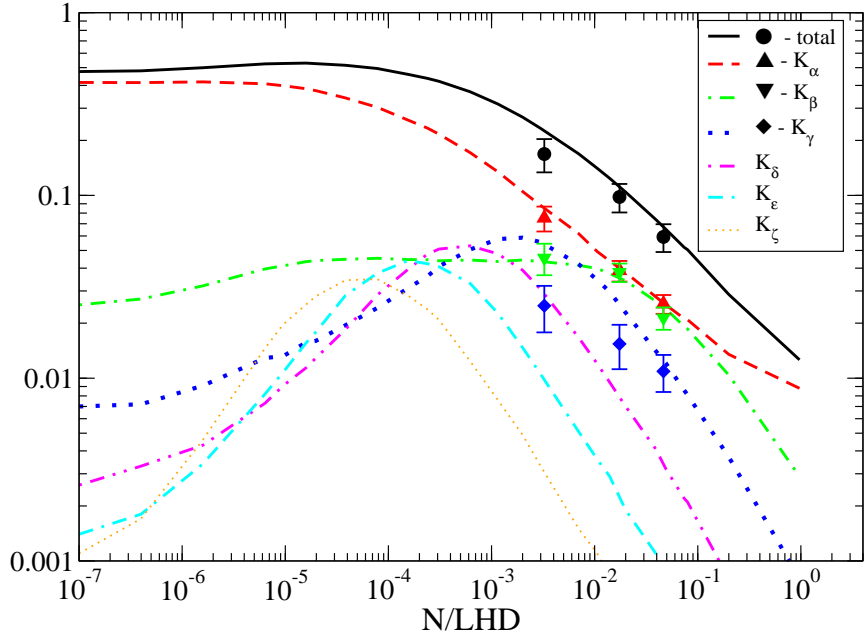}}
     \caption{The density dependence of the absolute $X$-ray yields in $\pi^- p$ atom.
     The experimental data are from \cite{19}.}
     \end{figure}

             The kinetic energy distribution of the pionic hydrogen at the
            moment of nuclear absorption in liquid hydrogen is
            shown in Figure 6. The experimental data from PSI experiments~\cite{7} (Bad.
            and the present theoretical description
            are shown on the left side and right side of the Figure~7 respectively.
             The most promising features are that for the first time we 
            can explain in more details  both the whole kinetic energy
            distribution of the $\pi^- p$ atoms at the instant of
            charge-exchange reaction and fine structure
            of this distribution. In particular, according to our calculations the component
            around $\sim 105$~eV and $\sim 209$~eV are explained by the $5\to 3$
	    and $3\to 2$ 
            Coulomb transitions, respectively. Besides, the relative
            yields of high energy components at the energies $\thicksim 105$
            eV and $\thicksim 209$ eV are in good agreement with
            the experimental data \cite{7}. Thus in contrast to previous
            theoretical studies the prediction of the present cascade model
            are in good agreement with the neutron time-of-flight spectrum at
            LHD.\\
     \begin{figure}[h!]
     \centerline{
     \includegraphics[width=0.44\textwidth,keepaspectratio]{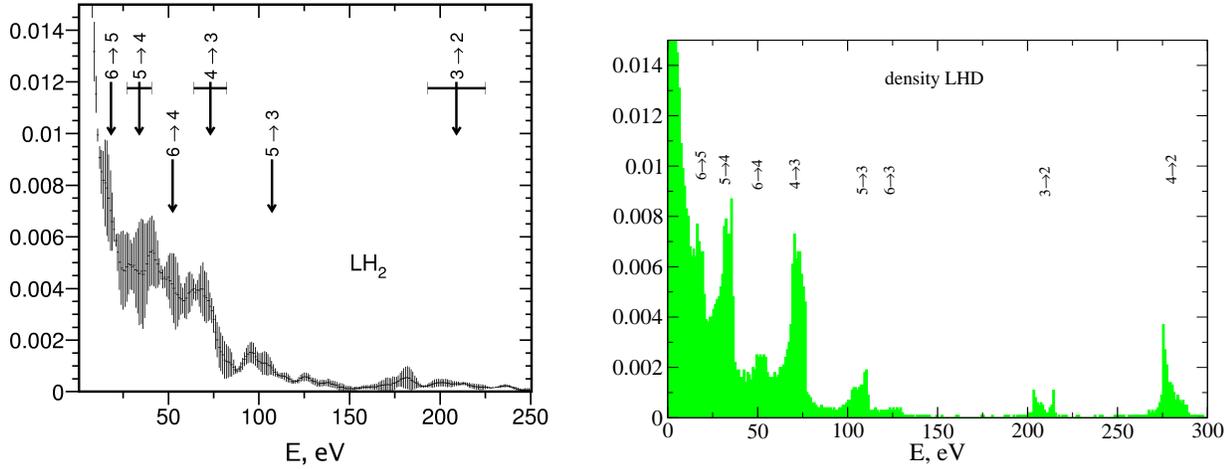}
     \hfill \includegraphics[width=0.5\textwidth,keepaspectratio]{ntof.eps}}
     \caption{The kinetic energy distribution of $\pi^- p$ at the instant of the nuclear
     absorption in liquid hydrogen: the experimental data \cite{7} ( on the left)
     in comparison with the results of the present cascade
     calculations (on the right)}
     \end{figure}

            In order to
            illustrate both the importance of the CD transitions with
            $\Delta n > 1$ and strength of the CD transitions with
            $\Delta n = 1$ the three variants of the cascade
            calculations were performed: full calculation where all
            CD transitions are taken into account, the calculation
            with only  $\Delta n = 1$ CD transitions and, finally,
            the calculation with  $\Delta n = 1$ CD transitions
            decreased by factor $k_{CD}=0.1$. The results introduced
            in Figure 7 show that in contrast to the CC approach \cite{12,13}
            the previous theoretical description of the CD process in
            the semiclassical \cite{20} and advanced
            adiabatic approaches (see \cite{21} and references therein) are
            in disagreement with the experimental data \cite{7}.

     \begin{figure}[h!]
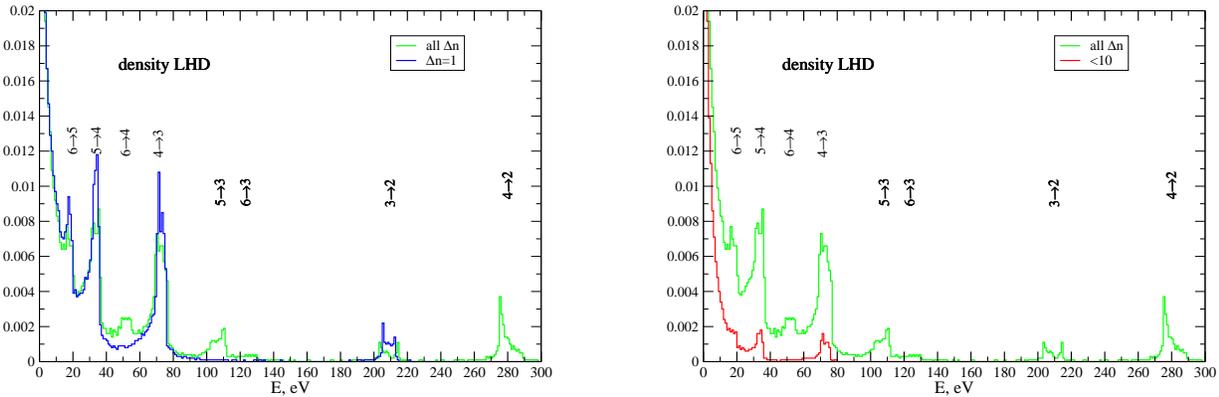

     \centerline{
     \includegraphics[width=0.45\textwidth,keepaspectratio]{ntof_dn_1.eps}
     \hfill \includegraphics[width=0.45\textwidth,keepaspectratio]{ntof_dn_1_01.eps}}
     \caption{The kinetic energy distribution of $\pi^- p$ at the instant of the nuclear
     absorption in liquid hydrogen for the different variants of the Coulomb transitions:
     full calculation  - green line and only $\Delta n = 1$
     CD transitions - blue line (on the left side); full calculation - green line and
     only $\Delta n = 1$ with the  factor $k_{Cd} = 0.1$ - blue line (on the right side)}
     \end{figure}

\section{Conclusion}
The present version of the cascade model significantly improves
the previous cascade calculations due to the self-consistent
treatment of the elastic scattering, Stark transitions and Coulomb
de-excitation. The inclusion in the cascade code the results of
the detailed quantum mechanical calculations of the differential
and integral cross sections for $nl\mapsto n'l'$ collisional
transitions allowed us to describe the kinetics of the atomic
cascade without employing any fitting parameters and more
accurately.\\
The predictions of this cascade model are in a good agreement with
the experimental data both for muonic and pionic hydrogen atoms.
In particularly, the  present cascade calculations allow for the
fist time to explain the observed kinetic energy distribution of
$\pi^- p$ atoms at the instant of nuclear reaction. Our results
explain the high energy components around $\sim 105$ eV and $\sim
209$ eV (due to $5\rightarrow 3$ and $3\rightarrow2$ Coulomb
transitions) and lead to a very good agreement with the
experimental weights of these components.

\section{Acknowledgements}
The authors would like to thank L.~Simons who initiated this
investigation as well as D.~Gotta, F.~Kottmann, V.~Markushin and
D.~Taqqu for many fruitful and stimulating discussions. This work
was supported by Russian Foundation for Basic Research Grant No.
$06-02-17156$.


\begin{thebibliography}{99}\itemsep -1mm
\bibitem{1} D.~Gotta, Prog.Part.Nucl.Phys. {\bf 52}, 133 (2004).
\bibitem{2} M.~Leon, H.A.~Bethe, Phys.Rev. {\bf 127}, 636 (1962).
\bibitem{3} E.~Borie, M.~Leon, Phys.Rev. A {\bf 21}, 1460 (1980).
\bibitem{4} T.P.~Terada, R.S.~Hayano, Phys.Rev. C {\bf 55}, 73 (1997)73.
\bibitem{5} J.F.~Crawford {\em et al}., Phys.Rev. D {\bf 43}, 46 (1991).
\bibitem{6} D.J.~Abbot {\em et al}., Phys.Rev. A {\bf 55}, 165 (1997).
\bibitem{7} A.~Badertscher {\em et al}., Europhys. Lett. {\bf 54}, 313 (2001).
\bibitem{8} F.~Kottman {\em et al}., Hyperf.Interact. {\bf 119}, 3 (1999);
 138(2001)55.
\bibitem{9} R.~Pohl {\em et al}., Hyperf. Interact. {\bf 138}, 35 (2001).
\bibitem{10} V.E.~Markushin, Phys.Rev. A {\bf 50}, 1137 (1994).
\bibitem{11} T.S.~Jensen, V.E.~Markushin, Eur.Phys. J. D {\bf 21}, 271 (2002).
\bibitem{12} G.Ya.~Korenman, V.N.~Pomerantsev, and V.P.~Popov, JETP. Lett.
     {\bf 81}, 543 (2005).
\bibitem{13} V.N.~Pomerantsev, V.P.~Popov,, JETP. Lett.
     {\bf 83}, 331 (2006); Phys. Rev. A {\bf 73}, 040501-1 (2006).
\bibitem{14} G.Ya.~Korenman, Hyperf. Interact. {\bf 101/102}, 81 (1996).
\bibitem{15} J.S.~Cohen, Rep. Prog. Phys. {\bf 67}, 1769 (2004).
\bibitem{16} H.~Anderhub {\em et al}., Phys. Lett. B {\bf 143}, 65 (1984).
\bibitem{17} N.~Bregant {\em et al}., Phys. Lett. A {\bf 241}, 344 (1998).
\bibitem{18} B.~Lauss {\em et al}., Phys. Rev. Lett. {\bf 80}, 3041 (1998).
\bibitem{19} A.J.~Rusi El Hassani {\em et al}., Z. Phys. A {\bf 135}, 113 (1995).
\bibitem{20} L.~Bracci, G.~Fiorentini, Nuovo Cimento A {\bf 43}, 9 (1978).
\bibitem{21} A.V.~Kravtsov, A.I.~Mikhailov, Yad. Fiz. {\bf 69}, 395 (2006).
\end{thebibliography}
\end{document}